\documentclass[%
 reprint,
 amsmath,amssymb,
 aps,
]{revtex4-2}

\usepackage{graphicx}
\usepackage{dcolumn}
\usepackage{bm}
\usepackage{amsmath}
\usepackage{braket} 
\usepackage{mathrsfs}

\begin{document}

\preprint{APS/123-QED}

\title{Apodized photonic crystals: A non-dissipative system hosting multiple exceptional points}

\author{Abhishek Mondal, Shailja Sharma}
\author{Ritwick Das}%
 \email{ritwick.das@niser.ac.in}
\affiliation{School of Physical Sciences, National Institute of Science Education and Research, An OCC of Homi Bhabha National Institute, Jatni - 752050, Odisha, India}%

\date{\today}

\begin{abstract}
Optical systems obeying non-Hermitian dynamics have been the subject of intense and concerted investigation over the last two decades owing to their broad implications in photonics, acoustics, electronics as well as atomic physics. A vast majority of such investigations rely on a dissipative, balanced loss-gain system which introduces unavoidable noise and consequently, this limits the coherent control of propagation dynamics. Here, we show that an all-dielectric, non-dissipative photonic crystal (PC) could host, at least two exceptional points in its eigenvalue spectrum. By introducing optimum apodization in the PC architecture, namely 1D-APC, we show that such a configuration supports a spectrum of exceptional points which distinctly demarcates the $\mathcal{PT}$- symmetric region from the region where $\mathcal{PT}$-symmetry is broken in the parameter space. The analytical framework allows us to estimate the geometric phase of the reflected beam and derive the constraint that governs the excitation of topologically-protected optical Tamm-plasmon modes in 1D-APCs.

\end{abstract}

\maketitle


\section{\label{sec:level1} INTRODUCTION}
Optical systems which are governed by non-Hermitian Hamiltonian dynamics through an engineered gain and dissipation mechanism, provide a route to overcome the limitations imposed by closed optical systems that obey the Hermitian-Hamiltonian led dynamics. Such non-Hermitian systems give rise to a real eigenvalue spectrum when the Hamiltonian commutes with the parity-time ($\mathcal{PT}$) operator. A continuous change in the parameter governing the Hermiticity (of the Hamiltonian) breaks the $\mathcal{PT}$ symmetry which manifests in the form of complex eigenvalues for the system. In the phase space, such points where the real and complex eigenvalues coalesce are termed as exceptional points (EPs) \cite{berry, Heiss_2012}. This spontaneous $\mathcal{PT}$-symmetry breaking has catalyzed a plethora of non-intuitive outcomes such as directional invisibility \cite{Zhu:14, PhysRevLett.106.213901}, coherent perfect lasing and absorption \cite{wan, PhysRevLett.105.053901, PhysRevA.82.031801, PhysRevLett.106.093902, PhysRevLett.112.143903}, negative refraction \cite{PhysRevLett.113.023903}, single-particle based sensing \cite{PhysRevA.93.033809, articleChen, PhysRevLett.112.203901}, distortion-free wireless optical power transfer \cite{articleXu} and a few more \cite{PhysRevA.85.023802, Doppler, Ota, PhysRevX.4.031011, PhysRevX.4.031042}. It is, however, worth noting that the incommensurate gain and loss distribution in non-Hermitian systems impose the primary limitation on the practical applications due to unpredictable signal-to-noise ratio near EP \cite{PhysRevB.92.235310, PhysRevLett.120.013901, PhysRevResearch.4.023009, PhysRevResearch.2.032057}. In order to circumvent such bottlenecks, a few possibilities have been explored. One such promising route is to create an asymmetric loss in the system (without gain) whose dynamics could be explored using a non-Hermitian Hamiltonian with a uniform background loss \cite{PhysRevB.92.235310, PhysRevLett.103.093902, PMID:25324384}. Such a configuration would exhibit $\mathcal{PT}$-symmetry which could be broken through scaling up the loss asymmetry. In a different scheme, a pseudo-Hermitian system was explored which allowed strong coupling between a large number of modes via manipulation of the parameters governing the Hamiltonian \cite{PhysRevLett.103.093902}. This led to the existence of EPs of multiple order and the interaction of eigenvalues around each EP provides a robust control on the propagation dynamics \cite{PhysRevLett.86.787, PhysRevA.85.064103}. In spite of the aforementioned developments, a useful and practical proposition would be to devise a configuration hosting a multitude of EPs with the constraint that the electromagnetic ($EM$) energy lost due to the non-Hermitian dynamics is stored in a reservoir. This essentially implies that the dissipative channel associated with a non-Hermitian system drives a separate Hermitian system which could allow reverse flow of $EM$ energy by virtue of cyclical dynamics. Such systems have been explored in the area of parametric frequency conversion processes where the $EM$ energy lost in one of the parametric processes (obeying non-Hermitian dynamics) is coherently added to the other parametric process that follows a Hermitian dynamics \cite{PhysRevLett.129.153901}. A plausible translation of such an idea in the non-absorptive linear systems would be to introduce a \emph{virtual} loss in an intermodal interaction process thereby generating multiple EPs in the parameter space. One of the simplest configurations imitating such a process is a multimodal interaction in an all-dielectric one-dimensional (1D) photonic-crystal (PC) with a gradually varying duty cycle (for each unit cell). In such an apodized 1D-PC, the forward (source) to backward (sink) mode-coupling dynamics is essentially governed by a pseudo-Hermitian Hamiltonian whose Hermiticity is determined by the apodization along the propagation direction. In the present work, we show the existence of multiple EPs in an apodized 1D-PC and develop an analytical framework for ascertaining the possibility of exciting topologically-protected optical edge modes in such aperiodically stratified configurations.   
\section{\label{sec:level1} Theoretical framework and coupled-mode formalism}
We consider a 1D-PC comprised of periodic bilayers with refractive indices $n_1$ and $n_2$ with thicknesses $d_1$ and $d_2$. Such conventional 1D-PCs or alternatively, distributed Bragg reflectors (DBRs) are usually characterized by photonic bandgaps (PBGs) which are separated from each other by high transmission (or pass) bands. In order to appreciate the $EM$ wave propagation dynamics, we consider the coupling between $p^{th}$-mode ($\ket{p}$) with $q^{th}$-mode ($\ket{q}$) which could be represented employing the coupled-amplitude equations given by \cite{yariv}
\begin{equation}\label{eq:equation1}
\frac{d A_{q}}{dz} = -i \frac{\beta_{q}}{|\beta_{q}|} \sum_{p} \sum_m {\tilde{\kappa}_{qp}}^{(m)} A_{p} e^{-i(\beta_{q} - \beta_{p} -m\frac{2\pi}{\Lambda}) z}
\end{equation}
where $\beta_p$ and $\beta_q$ are the longitudinal ($z$) components of wavevector $k_p$ and $k_q$ respectively. ${\tilde{\kappa}_{qp}}^{(m)}$ defines the strength of coupling (or coupling coefficient) between the $p^{th}$ and $q^{th}$ mode that is coupled through the $m^{th}$ Fourier component of the periodic dielectric distribution ( $\Lambda = d_1+d_2$). The factor $\Delta \beta = \beta_{q} - \beta_{p} -m\frac{2\pi}{\Lambda}$ (known as the phase-mismatch) is one of the dynamical variables (along with $\kappa_{qp}$) which dictate the measure of optical power transferred from one mode to the other. For the present work, we consider a contra-directional coupling set-up where a forward (along $+z$) propagating mode ($\ket{p} \equiv \ket{f}$) is coupled to a backward (along $-z$) propagating mode ($\ket{q} \equiv \ket{b}$). Accordingly, it could asserted that $\beta_b = - \beta_f$ or alternatively $\Delta\beta = 2\beta_f-\frac{2\pi}{\Lambda}$ and therefore, Eq. \eqref{eq:equation1} could be simplified to \cite{yariv}
\begin{equation} \label{eq:equation2}
  \frac{dA_b}{dz} =  i {\tilde{\kappa}} A_f e^{-i \Delta \beta z}  
\end{equation}
\begin{equation} \label{eq:equation3}
  \frac{dA_f}{dz} = - i {\tilde{\kappa}^{*}} A_b e^{i \Delta \beta z}  
\end{equation}
where $\tilde{\kappa} ~=~ \frac{i(1-\cos{2\pi \zeta})}{2\lambda} \frac{({n_1}^2 - {n_2}^2)}{\Bar{n}} = i\kappa$ and $\zeta$ is the dielectric filling fraction of layer with refractive index $n_1$ in the unit cell \emph{i.e.} $\frac{d_1}{\Lambda}$. The mean refractive index for an unit cell of thickness $\Lambda$ is $\Bar{n} ~=~ \sqrt{\frac{d_{1}{n_1}^2 +  d_{2}{n_2}^2}{\Lambda}}$. By using a Gauge transformation given by $[A_f, A_b] \rightarrow [\tilde{A}_f, \tilde{A}_b]e^{i/2[\Delta\beta_0 z - {\int_0}^z q(z') dz']}$, we obtain \cite{Sharma:21}
\begin{equation} \label{eq:equation4}
  i\frac{d}{dz} \begin{pmatrix} \tilde{A}_b\\ \tilde{A}_f\\ \end{pmatrix} = \begin{pmatrix} -\Delta k & -\tilde{\kappa}\\
\tilde{\kappa}^{*} & \Delta k \\ \end{pmatrix}
\begin{pmatrix} \tilde{A}_b\\ \tilde{A}_f\\ \end{pmatrix}
\end{equation}
Equation \eqref{eq:equation4} is analogous to time-dependent Schrödinger's equation with $t$-coordinate being replaced by $z$-coordinate. Here, $\Delta k$ ($=\frac{\Delta \beta}{2}$) and $q(z)=0$ remains constant (for a given frequency) across the 1D-PC which has a fixed duty cycle. The autonomous Hamiltonian $\hat{H} = -\vec{\sigma} \cdot \vec{B}$ with
$\vec{\sigma} \equiv [\sigma_x, \sigma_y, \sigma_z]$ are the Pauli's spin matrices and $\vec{B} \equiv [0, \kappa, \Delta k]$ (magnetic field analog) represents a pseudo-Hermitian evolution dynamics. In order to appreciate this point, we note that the eigenvalues of $\hat{H}$ which are given by $\emph{e}_{1,2} ~=~ \pm \sqrt{{\Delta k}^2 - {\kappa}^2}$ whereas the eigenfunctions are $\ket{\psi_1} ~=~ \begin{pmatrix} -i\frac{(\Delta k + \sqrt{{\Delta k}^2 - \kappa^2})}{\kappa}\\ 1 \\ \end{pmatrix}$ and $\ket{\psi_2} ~=~ \begin{pmatrix} +i\frac{(-\Delta k + \sqrt{{\Delta k}^2 - \kappa^2})}{\kappa}\\ 1 \\ \end{pmatrix}$. Here, $\tilde{\kappa} ~=~ i \kappa$. A closer look into the eigenvectors reveals that the equality $\kappa ~=~ \pm \Delta k$ manifests as coalescing of eigenvectors accompanied by vanishing eigenvalues. Such points in parameter space where $\kappa$ equals $\pm \Delta k$ are termed as exceptional points (EPs) and they distinctly demarcate the regions exhibiting Hermitian ($\mathcal{PT}$-symmetric phase) and non-Hermitian ($\mathcal{PT}$-broken phase) dynamical evolution of states (or modes).  

\begin{figure}[htbp]
\centering
\includegraphics[width=\linewidth]{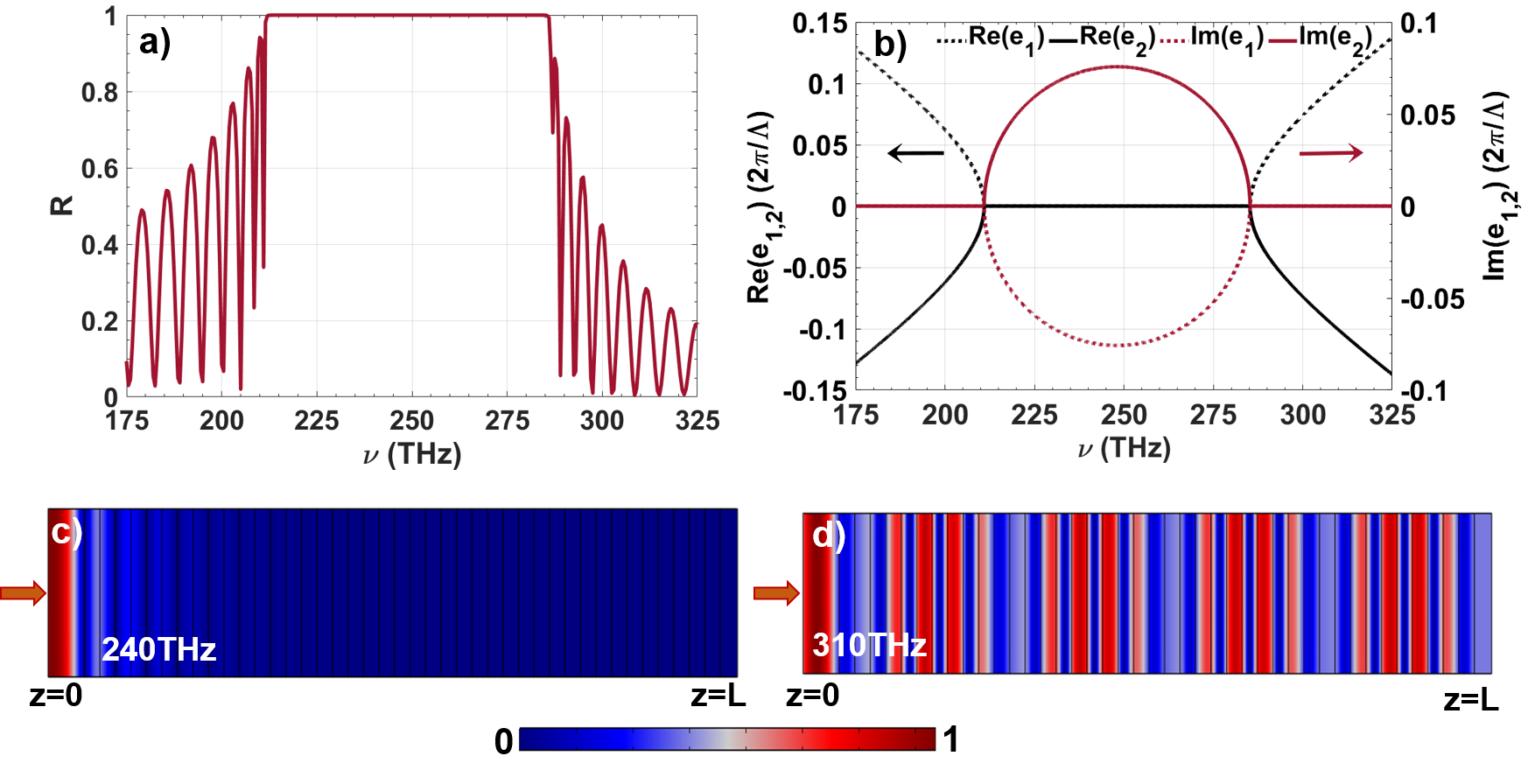}
\caption{a) Shows the reflection spectrum of a conventional (periodic) 1D-PC. b) Shows the variation in $Re(e_1)$ (dotted black curve), $Im(e_1)$ (dotted maroon curve), $Re(e_2)$ (solid black curve) and $Im(e_2)$ (solid maroon curve) as a function of frequency ($\nu$). c) and d) Shows the mode-field intensity for frequencies within the PBG $240~THz$ and that outside the PBG $310~THz$ respectively. The solid red arrow represents the direction of incidence of light.}
\label{Figure1}
\end{figure}

In order to appreciate the aforementioned idea, we consider a practical 1D-PC with $n_1~\equiv~TiO_2$ layer and $n_2~\equiv~SiO_2$ layer. The layer thicknesses are $d_1 = d_2 = 150~nm$. The reflection spectrum for $N = 20$ unit cells is plotted in Fig. \ref{Figure1}(a) which exhibits a high reflection band (or PBG) spreading over a $75~THz$ bandwidth. In order to obtain the reflection spectrum, finite element method (FEM) based simulations were carried out using the commercially available computational tool (COMSOL Multiphysics). In the simulations, the periodic boundary condition is imposed along the transverse direction and a mesh size of $5~nm$ is considered. We ignore the material dispersion for the simulations and assume $n_1 = 2.5$ ($\equiv TiO_2$) and $n_2 = 1.5$ ($\equiv SiO_2$) across the entire spectrum. For this 1D-PC, we also plotted the eigenvalues $e_1$ and $e_2$ (see Fig. \ref{Figure1}(b)) as a function of the frequency of the incident electromagnetic wave. It is apparent that the eigenvalues vanish at $\nu_1 \approx 210~THz$ and $\nu_2 \approx 285~THz$. These two frequencies ($\nu_1$ and $\nu_2$) define the EPs ($\kappa ~=~ +\Delta k$ and $\kappa ~=~ -\Delta k$) for the periodic 1D-PC. A closer look would also reveal that the eigenvalues are purely imaginary within the PBG and the band edges (Fig. \ref{Figure1} (a)) coincide with $\nu_1$ and $\nu_2$. The mode fields for frequencies lying inside the PBG $(240~THz)$ and  outside the PBG $(310~THz)$ are presented in Figs. \ref{Figure1}(c) and (d) respectively. 
It is worth noting that the investigations on systems exhibiting $\mathcal{PT}$-symmetry (or $\mathcal{PT}$-broken symmetry) led dynamics in photonics essentially involve optimally balanced gain-loss architectures such as segmented waveguides and photonic crystals. In such systems, a complex relative permittivity in different sections depicting \emph{actual} gain or loss for the propagating light beam gives rise to the $\mathcal{PT}$-symmetry (or $\mathcal{PT}$-broken symmetry). The present configuration involving 1D-PC does not include an \emph{actual} dissipative component for achieving the $\mathcal{PT}$-symmetric to $\mathcal{PT}$-symmetry broken phase transition. Alternatively, the coupling of optical power to the backscattered mode $\ket{b}$ is analogous to a \emph{virtual} loss for a forward propagating $\ket{f}$ mode. When this coupling is relatively weak \emph{i.e.} $\Delta k > \tilde{\kappa}$, $\ket{f}$ and $\ket{b}$ exhibits cyclic exchange of optical power (as a function of $z$) which is a primitive outcome for a $\mathcal{PT}$-symmetric dynamics. On the other hand, a strong coupling regime where $\Delta k < \tilde{\kappa}$ manifests through a monotonic growth of backscattered mode ($\ket{b}$) that is a signature of $\mathcal{PT}$-symmetry broken phase. It is worthwhile to reiterate the point that the two regimes depicted by the inequality of $\Delta k$ and $\tilde{\kappa}$ (in the parameter space) could be mapped onto the PBG and pass or transmission band (s) in the reflected spectrum. Subsequently, each PBG is necessarily bounded by two EPs in this framework. Additionally, these two EPs are fixed and could not be tailored for a given 1D-PC with a fixed duty cycle and fixed period. Also, the conventional 1D-PC geometry excludes the possibility of realizing higher-order exceptional points \cite{PhysRevA.101.063829}. Taking a cue from this critical viewpoint, we note that a small apodization or gradual change in dielectric filling fraction ($\zeta$) of each unit cell of the 1D-PC would allow us to realize discretely spaced (multiple) EPs at different optical frequencies (or wavelengths). In order to elucidate this point, we recall that $\Delta k$ as well as $\tilde{\kappa}$ is a function of $\zeta$. An optimum spatial variation in $\zeta$ could essentially give rise to the possibility of EPs at different physical locations (along $z$) in a 1D-PC. As an example, we show below that an optimally apodized 1D-PC (1D-APC) which satisfies the adiabatic constraints enables us to observe EPs at discreetly separated points along $z$.  
\begin{figure}[htbp]
\centering
\includegraphics[width=\linewidth]{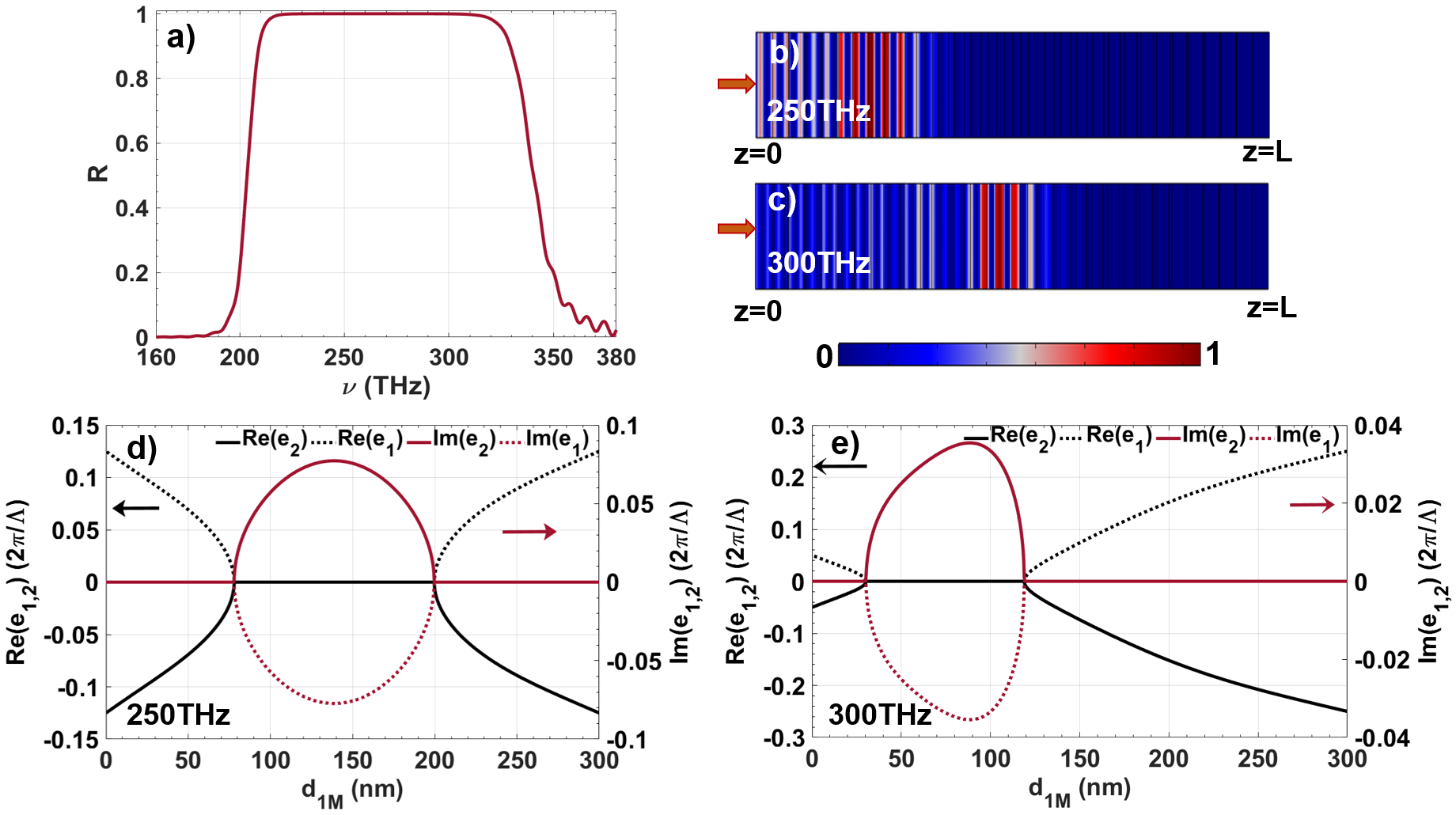}
\caption{a) Shows the reflection spectrum for designed 1D-APC. (b) and (c) Shows the mode-field intensities for two different frequencies $\nu_a = 250~THz$ and $\nu_b =300~THz$ which are within the PBG of 1D-APC. (d) and (e) Shows the variation in $Re(e_1)$ (dotted black curve), $Im(e_1)$ (dotted maroon curve), $Re(e_2)$ (solid black curve) and $Re(e_2)$ (solid maroon curve) as a function of $TiO_2$ layer thickness for each unit cell (\emph{i.e.} $d_{1M}$) at frequencies $\nu_a = 250~THz$ and $\nu_b = 300~THz$ respectively.}
\label{Figure2}
\end{figure}
\subsection{\label{sec:level2} Design of an 1D apodized PC and intermodal coupling} 

We consider a 1D-PC configuration that exhibits varying dielectric filling fraction ($\zeta$) in each unit cell. This variation is essentially dictated through the relation $d_{1M} = d_1- M\delta$ and $d_{2M} = \Lambda - d_{1M}$. Here, $d_{1M}$ and $d_{2M}$ are the thickness of $TiO_2$ and $SiO_2$ layers respectively in $M^{th}$ unit cell ($M = 0,1,2,3,...,(N-1)$ for $N$ number of unit cells). The unit cell period, however remains unchanged \emph{i.e.} $\Lambda = d_{1M}+d_{2M} = d_1+d_2$. This apodization in 1D-PC could be visualized through a longitudinal variation in $\Delta k$ as well as $\tilde{\kappa}$ by virtue of a monotonic change in average refractive index ($\bar{n}$) for an \emph{unit cell}. This variation in $\Delta k$ and $\tilde{\kappa}$ in a 1D-APC geometry leads to an adiabatic evolution of the Stokes vector along the propagation direction and manifests through a broader PBG ($\approx 140~THz$) in comparison with a conventional (periodic) 1D-PC \cite{Sharma:21}. This is presented in Fig. \ref{Figure2}(a) which shows a broader reflection spectrum for the 1D-APC in comparison with the conventional 1D-PC (Fig.\ref{Figure1}(a)). In addition, a flat transmission band and the absence of sharp transmission resonances is a distinct feature of 1D-APC. The mode-propagation characteristics for the frequencies within the PBG (of 1D-APC) is explored by drawing a comparison with the mode-field distributions for the equivalent modes within the PBG of a conventional 1D-PC. Figures \ref{Figure2}(b) and (c) shows the mode-field distribution for two frequencies $\nu_a = 250~THz$ and $\nu_b = 300~THz$ which are within the PBG of 1D-APC. In comparison with the mode-field distribution shown in Fig. \ref{Figure1}(c), it could be observed that different modes are reflected from spatially separated $z$ values. The smaller frequency ($\nu_a = 250~THz$) is reflected from the regions which are closer to $z = 0$ edge of the 1D-APC in comparison to that for $\nu_b = 300~THz$. This variation is indicative of the fact that the field is localized and exhibits instantaneous localization in different 1D-APC sections.  
\begin{figure}[htbp]
\centering
\includegraphics[width=\linewidth]{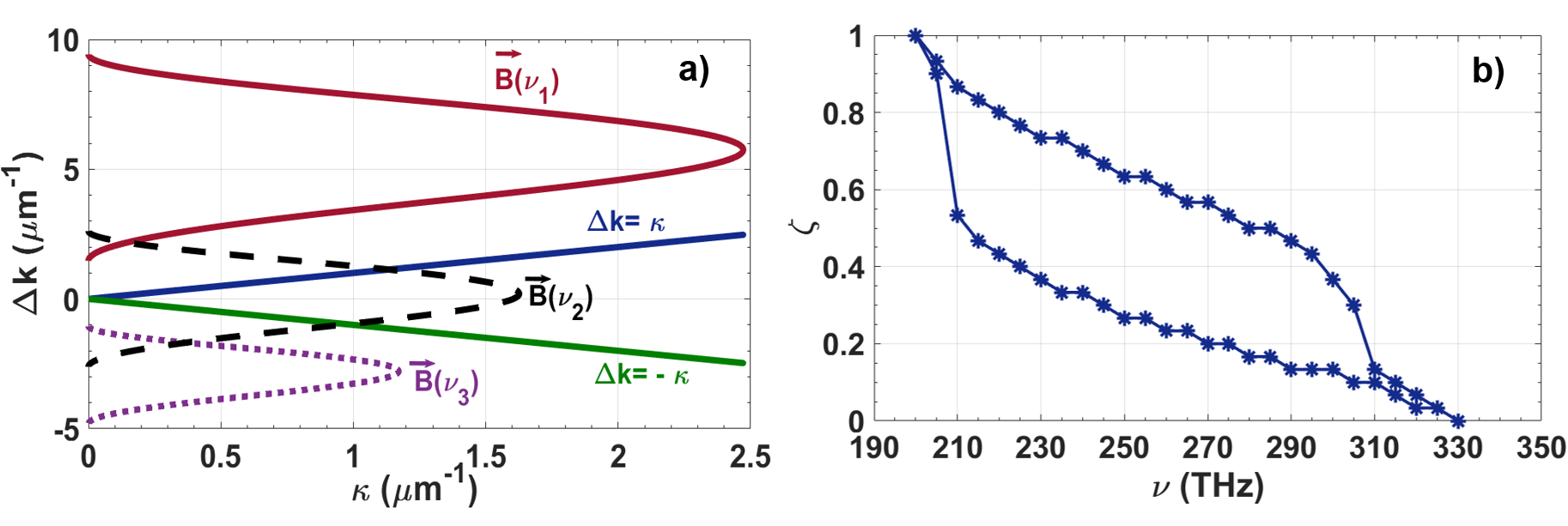}
\caption{(a) Shows the variation of $\vec{B}$ in parameter space (spanned by $\kappa$ and $\Delta k$) at different operating frequencies ($\nu_1 ~=~ 400~THz$, $\nu_2 ~=~ 250~THz$, $\nu_3 ~=~ 160~THz$) for the designed 1D-APC. The blue and green solid lines represent the $\Delta k = \kappa$ and $\Delta k = -\kappa$ curves. (b) Shows the location of EPs in different unit cells (with different filling fraction $\zeta$) as a function of frequency ($\nu$).}
\label{Figure3}
\end{figure}
From a different perspective, it is apparent that the variation in dielectric filling fraction ($\zeta$) would result in different eigenvalues (and corresponding eigenvectors) for each unit cell. Accordingly, we plot the  eigenvalues $e_1$ and $e_2$ as a function of $d_{1M}$ for two frequencies $\nu_a = 250~THz$ (Fig. \ref{Figure2}(d)) and $\nu_b = 300~THz$ (Fig. \ref{Figure2}(e)) which are within the PBG of 1D-APC. Each one of the figures shows that the eigenvalues ($e_1$ and $e_2$) vanish at two different values of $d_{1M}$ \emph{i.e.} at the location of two different unit cells. Therefore, the 1D-APC geometry hosts two EPs for every $d_{1M}$. Consequently, for a multitude of $\zeta$, there would be multiple EPs in the 1D-APC for a forward-propagating mode to a backscattered mode-coupling process. As discussed before, the regions where $\Re{e_1}$ and $\Re{e_2}$ are non-zero in Figs. \ref{Figure2}(d) and \ref{Figure2}(e) exhibit a $\mathcal{PT}$-symmetric coupling dynamics between the forward-propagating and backscattered modes. On the other hand, in the regions where $e_1$ and $e_2$ are purely imaginary, the mode-coupling process exhibits $\mathcal{PT}$-symmetry broken manifolds. The illustrations presented in Figs. \ref{Figure2}(d) and \ref{Figure2}(e) show that for each frequency within the PBG, the 1D-APC hosts two EPs at two different $d_{1M}$. This essentially implies that there exists one or more than one EPs hosted by each unit cell of the 1D-APC. Therefore, an 1D-APC is expected to host multiple EPs which are spectrally as well as spatially separated from each other. 
In order ascertain the spectral location of EPs in the 1D-APC, we plot the evolution of $\vec{B}$ in the parameter space for three different frequencies $\nu_1 ~=~ 400~THz$, $\nu_2 ~=~ 250~THz$, and $\nu_3 ~=~ 160~THz$ as shown in Fig.\ref{Figure3}(a). It could be noted at $\nu_1$ and $\nu_3$ are situated outside PBG of 1D-APC (see Fig. \ref{Figure2}(a)). Since, the EPs are depicted by the condition $\Delta k = \vert\kappa\vert$, Fig.\ref{Figure3}(a) also contains the curve $\Delta k = \pm\kappa$ (solid blue and green curves). It is apparent that $\Delta k = \pm\kappa$ curve intersects $\vec{B}_{\nu_2}$ at two points and it does not intersect the $\vec{B}_{\nu_1}$ curve as well as the $\vec{B}_{\nu_3}$ curve in the parameter space. For frequencies close to the band-edge of 1D-APC (say $200~THz$ or $350~THz$), it could be ascertained that there exists only one EP in the eigenvalue spectrum. This is primarily due to the adiabatic constraints followed by the 1D-APC design. In other words, for the band-edge frequencies, the forward and backward propagating modes are decoupled ($\tilde{\kappa}$) at entry ($z = 0$) and exit ($z=L$) face of the crystal. Additionally, $d_{1M} = \Lambda $ for $m = 0$ (or $d_{2M} = \Lambda $ for $m = N$) in case of band-edge frequencies that leads to $\Delta k = 0$ for $\zeta =$ (or $\zeta = 1$). Therefore, $\tilde{\kappa} = \Delta k = 0$ depicts the only EP for the band-edge frequencies.\\ 

In order to elucidate the aforementioned point, we present the spectral location of EPs as a function of dielectric filling fraction ($\zeta$) or propagation direction ($z$) in Fig. \ref{Figure3}(b). It could be observed that there exists two (2) EPs (at different $\zeta$ or $z$) for all the frequencies well within the PBG of 1D-APC. However, for the band-edge frequencies ($\nu_l = 200~THz$ and $\nu_h = 330~THz$), the 1D-APC hosts one EP only. Nevertheless, the area enclosed by the EPs in Fig. \ref{Figure3}(b) represents the region of $\mathcal{PT}$-symmetry broken phase for the 1D-APC. It is interesting to note that the separation between the two EPs for frequencies closer to the band-edges (say $\nu \leq 210~THz $ or $\nu \geq 310~THz$) very less and they tend to overlap at the same filling fraction. It is important to note that these EPs are physically positioned \emph{close to} the entry ($z=0$) and exit ($z=L$) face of the 1D-APC where $\tilde{\kappa}$ is very small. By virtue of this, the PBG corresponding to that unit cell of 1D-APC is relatively smaller in comparison with the PBG for a unit cell close to the center ($z \approx\frac{L}{2}$) of 1D-APC. Due to the fact that the EPs exist at the band-edges of PBG for each unit cell of APC, a smaller PBG would essentially imply closely spaced EPs near the band-edges (see Fig. \ref{Figure3}(b)).     



\subsection{\label{sec:level2} Geometric phase estimation of reflection band}  
\begin{figure}[htbp]
\centering
\includegraphics[width=\linewidth]{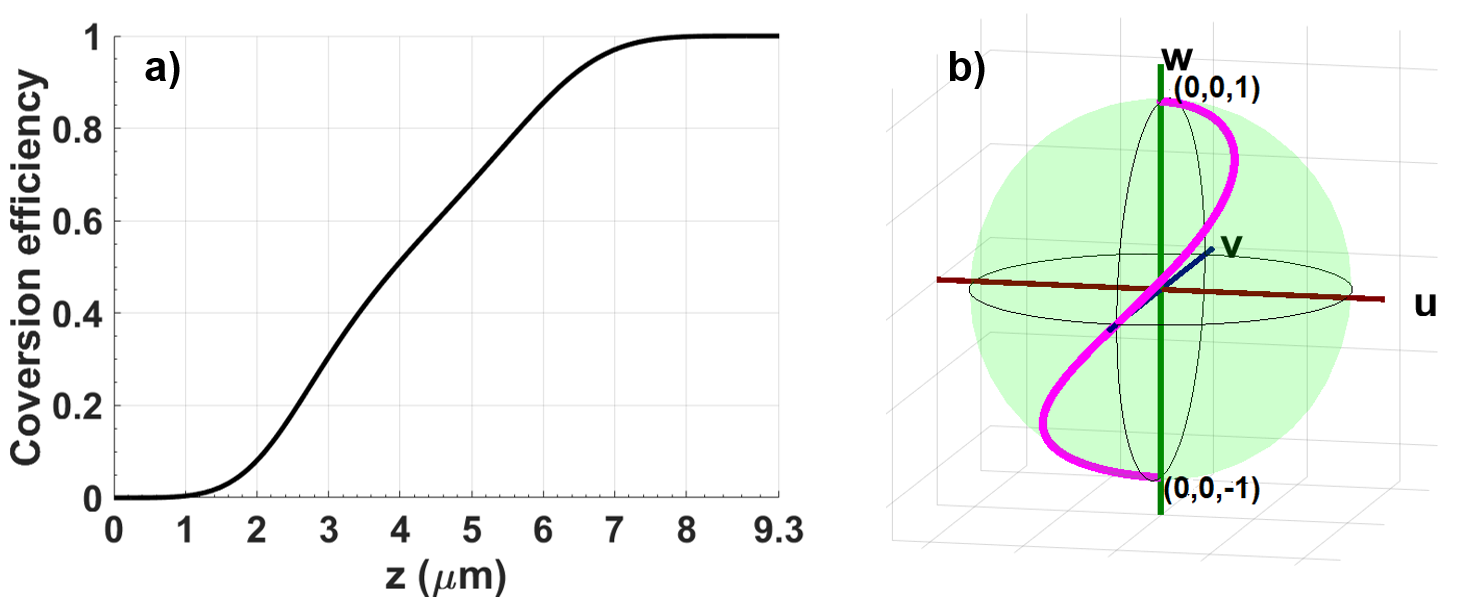}
\caption{a) Shows the variation in conversion efficiency ($\frac{w+1}{2}$) for optical power transfer between a forward-propagating mode to a backscattered mode as a function of 1D-APC length ($z$) for a frequency $\nu_2 = 250~THz$ which is within the PBG. (b) Presents the state-vector ($\vec{S} = [u,~v,~w]$) trajectory on the Bloch sphere for $\nu_2~=~250~THz$. }
\label{Figure4}
\end{figure}

It is well known that the geometric phase of a pass-band (or transmission band) for a one-dimensional conventional photonic crystal is quantized ($0$ or $\pi$) and it is known as the `Zak' phase. However, the geometric interpretation of backscattered (or reflection) phase from a 1D-PC remains irrelevant. However, in case of 1D-APC, the reflection of different spectral components (within the PBG) takes place from different unit cells (or $z$) along the propagation direction \cite{Sharma:21}. For example, the adiabatic following constraint leads to conversion of optical power from the forward-propagating to the backscattered mode predominantly towards the exit face of 1D-APC for frequency $\nu = 250~THz$ which could be seen in Fig. \ref{Figure4}(a). Through a similar route, it could be shown that different spectral components within the PBG are reflected strongly from different unit cells of 1D-APC \cite{Sharma:21}. The primary underlying reason could be traced to the variation in $\tilde{\kappa}$ and $\Delta k$ for each spectral component in the PBG which are non-identical. Consequently, the estimation of geometric phase acquired by different backscattered modes is expected to be different and must play a crucial role in establishing the \emph{bulk-boundary} correspondence in case of 1D-APC. In order to obtain the geometric phase $\gamma$, we consider a triad defining the state vector $\vec{S}$ ($\equiv[u, v, w]$) where $u = \Tilde{A}_i {\Tilde{A}_r}^* + \Tilde{A}_r{\Tilde{A}_i}^*$, $v = -i[\Tilde{A}_i {\Tilde{A}_r}^* - \Tilde{A}_r {\Tilde{A}_i}^*]$ and $w = {| \Tilde{A}_r|}^2 - {| \Tilde{A}_i|}^2$ \cite{Sharma:21}. 
The $z$-component of the state-vector ($w$) represents the conversion efficiency of optical power from a forward-propagating to a backscattered mode \cite{Sharma:21}. It is also worth noting that the trajectory of state-vector ($\vec{S}$) corresponding to the frequencies within the PBG is non-closed. Alternatively, the geometric phase is not a conserved quantity during the dynamical evolution of states owing to the $\mathcal{PT}$-symmetry broken phase. In general, the solid angle subtended by the state-vector trajectory at the center of the Bloch sphere is used for computing the geometric phase. However, in case of an adiabatic evolution, the state-vector trajectory could be very complicated. In Fig. \ref{Figure4}(b), we have plotted such a state-vector trajectory (on the Bloch sphere) corresponding to a frequency $\nu = 250~THz$ (which is within the PBG of 1D-APC). It is important to note that $\vec{S} = [0,0,-1]$ and $\vec{S} = [0,0,1]$ represent states in which all the optical power ($\propto \vert \tilde{A}_{f,b} \vert ^2$) is present in the forward-propagating and backscattered mode respectively. Although, the adiabatic evolution of state-vector results in complete optical power transfer from the forward to backward-propagating mode \emph{i.e.} $w = -1$ to $w = 1$, the estimation of acquired geometric phase is quite complicated owing to the spiralling trajectory of $\vec{S}$ on the Bloch-sphere. However, it is interesting to note that $\vec{S}$ goes from [$0,0,-1$] to [$0,0,1$] for all the frequencies within the PBG of 1D-APC by virtue of satisfying the adiabatic following constraints. The most important point is to note that the conversion efficiency (or reflectivity) is `unity' for all the frequencies within the PBG of 1D-APC \cite{Sharma:21}. In other words, $\vec{B}$ goes from $[0,0,-\Delta k]$ to $[0,0,\Delta k]$ in the parameter space for all the PBG frequencies (through any trajectory) when the adiabatic following constraints are satisfied \cite{Sharma:21}. 

\begin{figure}[htbp]
\centering
\includegraphics[width=\linewidth]{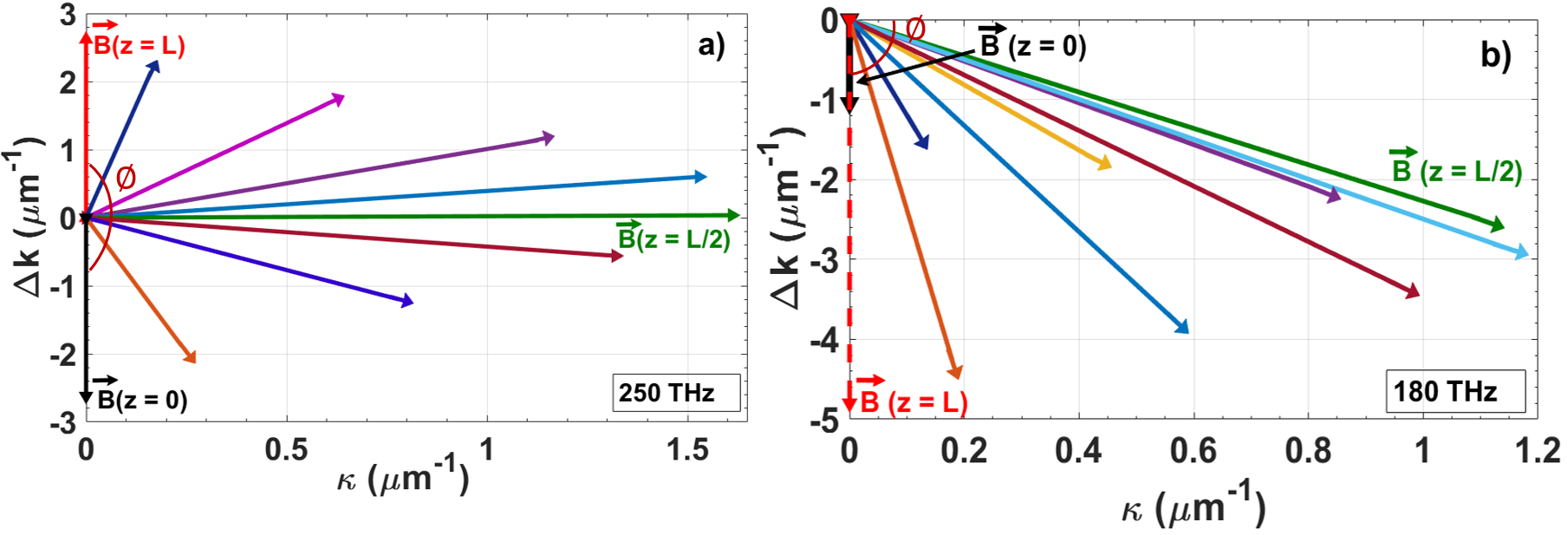}
\caption{Represents the evolution of $\vec{B}$ as a function of length ($L$) of 1D-APC in parameter ($\Delta k - \kappa$) space for a) $\nu_2 = 250~THz$ and b) $\nu_4 = 180~THz$. $\phi$ represents the angle subtended by curve $\vec{B}$ at the origin.}
\label{Figure5}
\end{figure}
By virtue of the fact that the state-vector $\vec{S}$ adiabatically follows $\vec{B}$ (as per the Bloch equation), the initial and the final value of $\vec{B}$ could also yield the geometric phase ($\gamma$). It is known that $\gamma$ is estimated from angle $\phi$ (subtended by $\vec{B}$ at the origin $\Delta k = \tilde{\kappa} = 0$) through the relation $\gamma = \frac{\phi}{2}$. In that case, the geometric phase for each spectral component within the PBG is $\frac{\pi}{2}$. In order to elucidate this point, we plot $\vec{B}$ at different $z$ of 1D-APC in the parameter space for $\nu = 250~THz$ as shown in Fig. 5(a). At the entry face of 1D-APC ($z = 0$), $\vec{B}(z=0) = [0,0,-2.7~\mu m^{-1}]$ (black arrow) and gradually goes to $\vec{B}(z=L) = [0,0,+2.7~\mu m^{-1}]$ (red arrow) at $z = L$. At $z = \frac{L}{2}$, $\Delta {k} = 0$ and $\tilde{\kappa}$ is maximum (green arrow in Fig. \ref{Figure5}(a)) The evolution of $\vec{B}$ in Fig. \ref{Figure5}(a) yields $\phi = \pi$ and consequently, $\gamma = \frac{\pi}{2}$. In a similar manner, $\gamma$ for all the frequencies within the PBG would be $\frac{\pi}{2}$ by virtue of adhering to the constraints imposed by adiabatic following. Hence, it could be asserted that a geometric phase of $\frac{\pi}{2}$ is acquired by a reflected beam in a 1D-APC for the values of parameters which results in $\mathcal{PT}$-symmetry broken phase. On the contrary, the variation in $\vec{B}$ is plotted as a function of $z$ for $\nu = 180~THz$ which is outside the PBG of 1D-APC (see Fig. \ref{Figure5}(b)). $\vec{B}(z =0)$ (black arrow) and $\vec{B}(z = L)$ (red dashed arrow) are both negative as well as co-parallel in this case. Consequently, the geometric phase $\gamma = \frac{\phi}{2} = 0$ for $\nu = 180~THz$. In addition, it is apparent that $\Delta k \neq 0$ at any point (or any $z$) in the 1D-APC.

\subsection{\label{sec:level2} Tamm-plasmon excitations in 1D-APC and topological connection} 
\begin{figure}[htbp]
\centering
\includegraphics[width=\linewidth]{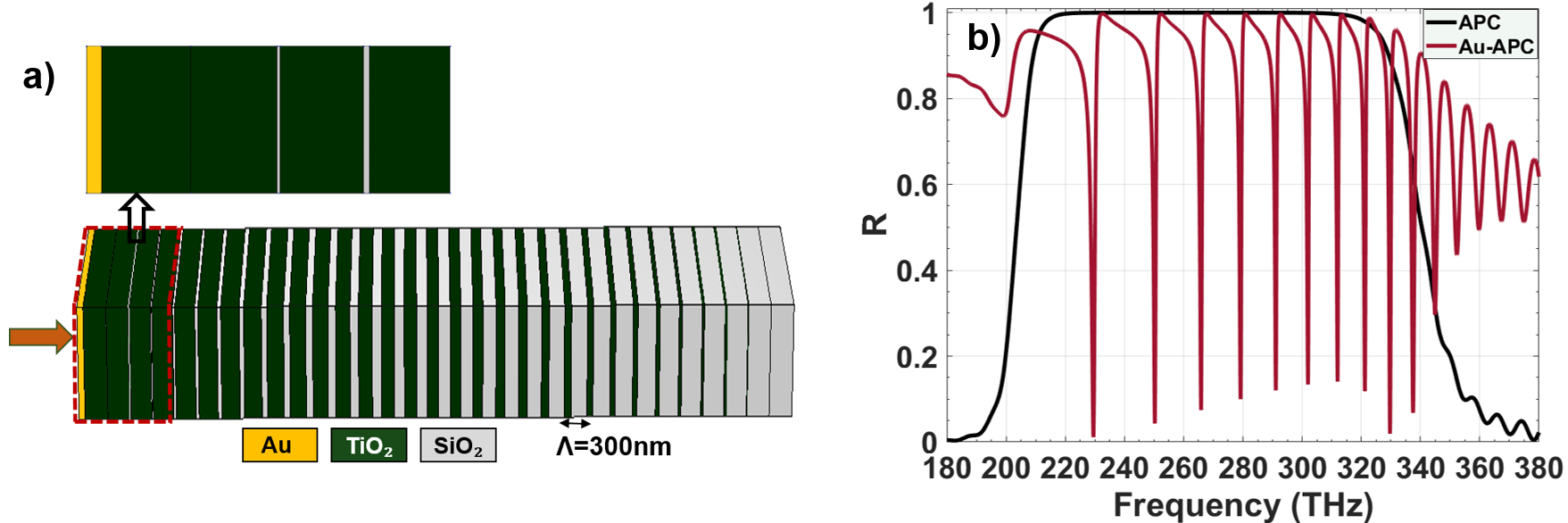}
\caption{a) Shows the schematic of the $Au$-1D-APC heterostructure. The $Au$-layer is placed adjacent to the high-index $TiO_2$ layer. The thick brown arrow depicts the direction of light incidence on the Au-1D-APC b) Shows the simulated reflection spectrum of 1D-APC without $Au$ (black solid curve) and that of $Au$-1D-APC (maroon solid curve).}
\label{Figure6}
\end{figure}

The presence of a plasmon-active layer adjacent to the all-dielectric 1D-APC results in excitation of multiple Tamm-plasmon modes which are non-degenerate. As an example, we consider a thin ($d_{Au} = 30~nm$) layer of gold placed in contact with high index layer ($TiO_2$) of 1D-APC (see Fig.\ref{Figure6}(a)). The simulated reflection spectrum (using transfer matrix method) exhibits a sharp resonance within the PBG as shown in Fig.\ref{Figure6}(b). These resonances are essentially due to Tamm-plasmon mode excitations which are highly localized electromagnetic states. Figure \ref{Figure6}(b) depicts the existence of $10$ Tamm-plasmon modes within the PBG of 1D-APC. Although there are a few sharp resonances outside the PBG, their mode-field signatures do not resemble that for a Tamm-plasmon mode \cite{Sharma}. In general, the existence of Tamm-plasmon modes is governed by the condition $\phi_{APC} + \phi_{Au}~=~ 2s\pi$ where $s~=~ 0,~1,~2,~3 ....$ is an integer \cite{Shukla:18, PhysRevX.4.021017, PhysRevB.74.045128}.
Here, $\phi_{APC}$ is the total phase acquired by the reflected beam from the 1D-APC (light incident from $Au$ side), and $\phi_{Au}$ is the phase acquired by reflected beam at the $Au-TiO_2$ interface. It is worthwhile to reiterate that the dielectric layer (of 1D-APC) adjacent to the $Au$-film is $TiO_2$ which is the high index layer. In the present context $\phi_{APC} = \gamma + \alpha$, where $\alpha$ is the dynamic phase acquired by the reflected beam \cite{Sharma:21}. This could be estimated by noting the fact that the EPs (for a given frequency) are situated in different unit cells (or $\zeta$) of the 1D-APC. For a frequency $\nu$, if the nearest EP (with respect to $z =0$) is present in the $p^{th}$-unit cell of 1D-APC, then $\alpha$ could be determined using
\begin{equation}\label{eq:equation5}
    \alpha = \frac{2\pi \nu}{c}\sum_{M = 0}^p [n_1 d_{1M} + n_2 d_{2M}]
\end{equation}

The knowledge of location for EPs in the 1D-APC (obtained from the eigenvalue spectrum of $\hat{H}$) would accurately yield the dynamic phase ($\alpha$) for any frequency of operation ($\nu$). In conjunction with the estimate of $\gamma$, this information would allow us to determine the Tamm-plasmon mode resonance frequencies ($\nu_r$). This recipe provides a flexibility in terms of designing an 1D-APC which would facilitate excitation of Tamm-plasmon mode at a target (desirable) frequency (or wavelength) of operation. One such application could be the generation of higher harmonics or frequency downconversion using optical surface states \cite{PhysRevB.97.115438}. In this case, the 1D-APC could be designed such that the Tamm-plasmon modes (localized modes) have resonance frequencies that are governed by the energy conservation and phase-matching constraints imposed by the frequency conversion process.     

\section{Conclusions} 
In conclusion, we presented an all-dielectric 1D-APC design which hosts multiple exceptional points in its eigenvalue spectrum by virtue of exhibiting a non-Hermitian dynamics for a mode-coupling process between a forward-propagating mode to its backscattered counterpart. Although, the 1D-APC does not include any dissipative component, the intermodal coupling mechanism could be classified in terms of $\mathcal{PT}$-symmetric and $\mathcal{PT}$-broken phases which are connected through a quantum phase-transition. We also showed that the reflected beam (within the PBG) acquires a geometric phase of $\frac{\pi}{2}$ in the $\mathcal{PT}$-symmetry broken phase. As a consequence of this outcome, the 1D-APC could be designed for exciting the optical Tamm-plasmon modes at any desirable frequency within the PBG. This design flexibility allows us to employ such architectures for quite a few applications such as efficiently carrying out optical frequency conversion using surface states \cite{PhysRevB.97.115438}.

\section{Disclosures}
The authors declare that there are no conflicts of interest related to this article.

\bibliography{ref}

\end{document}